\documentclass[10pt]{article}

\usepackage{epsfig}

\def\beq{\begin{equation}}  

\def\eeq{\end{equation}}

\newcommand{\ssim}{\sim\!}

\newcommand{\eq}{\begin{equation}}

\newcommand{\eqx}{\end{equation}}

\newcommand{\eqn}{\begin{eqnarray}}

\newcommand{\eqnx}{\end{eqnarray}}

\begin{document}
 
\title{Sensitivity to the Standard Model Higgs Boson\\ in Exclusive Double Diffraction}
\author{M. Boonekamp\thanks{CEA/DSM/DAPNIA/SPP, CE-Saclay, F-91191
Gif-sur-Yvette Cedex, France}, R. Peschanski\thanks{CEA/DSM/SPhT,  Unit\'e 
de recherche associ\'ee au CNRS, CE-Saclay, F-91191 Gif-sur-Yvette Cedex,
France}, C. Royon\thanks{CEA/DSM/DAPNIA/SPP, CE-Saclay, F-91191 Gif-sur-Yvette Cedex,
France}}
\maketitle

\begin{abstract}

We use a Monte Carlo implementation of recently developed models of double diffraction
to assess the sensitivity of the LHC experiments to Standard Model Higgs bosons produced in 
exclusive double diffraction. The signal is difficult to extract, due to experimental limitations related to
the first level trigger, and to contamination by inclusive double diffractive background.
Assuming these difficulties can be overcome, the expected signal-to-background ratio is 
presented as a function of the experimental resolution on the missing mass. With a missing mass 
resolution of 2 GeV, a signal-to-background ratio of about 0.5 is obtained; a resolution of 1 GeV 
brings a signal to background ratio of 1. This result is lower than previous estimates, and the discrepancy 
is explained.

\end{abstract}

\section{Introduction}

The subject of Higgs boson production in double diffraction (denoted DPE, for Double Pomeron Exchange) has drawn 
considerable interest in recent years \cite{others,bialas1,bialas2,us,cox,sci,khoze}.
Many approaches have been pursued, considering diffractive scattering in the Regge picture \cite{bialas1,bialas2,us,cox}, 
as final state soft color interactions \cite{sci}, or as fully perturbative exchange of gluon pairs \cite{khoze}.

One generally considers two types of DPE events, namely ``exclusive'' DPE, where the central heavy object is produced
alone, separated from the outgoing hadrons by rapidity gaps : 

\eq p p \rightarrow p + H + p, \eqx

\noindent and ``inclusive'' DPE, where the colliding Pomerons are resolved (very much like ordinary hadrons), dressing 
the central object with Pomeron ``remnants'':

\eq p p \rightarrow p + X + H + Y + p. \eqx

In general, exclusive Higgs boson production is considered most promising for both experimental and theoretical reasons
which will be recalled later on. Although a less appealing search channel, inclusive DPE 
is important to consider since it constitutes a background to exclusive DPE. Besides, it should not be forgotten
that of the above two, only inclusive DPE has actually been observed for high central masses~\cite{cdfdijet}. Exclusive DPE 
for masses exceeding about 4 GeV is still hypothetical. 

A recently developed Monte-Carlo program, {\tt DPEMC} \cite{dpemc}, proposes an implementation of the models of 
\cite{bialas1,bialas2,us,cox}. It uses {\tt HERWIG} \cite{herwig} as a cross-section library of hard QCD 
processes, and when required, convolutes them with the relevant Pomeron fluxes and parton densities. 

On the experimental side, performance simulations of a possible 
experimental setup for forward proton detection at the LHC are available \cite{helsinki}. The LHC experiments ATLAS and CMS
also propose tools for fast simulation of the response of their detectors \cite{cmsim}. All needed ingredients are thus 
present to allow for a consistent evaluation of the DPE Standard Model Higgs boson search potential, including experimental effects. 
Such a study has not been performed yet. We focus on the $H \rightarrow b\bar{b}$ final state, which dominates the 
cross-section in the mass range 100-140 GeV.

In the next section, the theoretical framework is recalled,
with some attention devoted to the exclusive processes. Relevant backgrounds are mentioned, and some details of the 
simulation are given. The following section describes the experimental context. The most important steps of the 
analysis are then given, concentrating on trigger aspects, background rejection, and mass reconstruction. The results are 
given as a function of the expected missing mass resolution. Conclusions follow.

We do not pretend to exhaust all possibilities in this paper, but give an idea of what can be achieved under reasonably
optimistic conditions. Further details and ideas for improvement will be given in a forthcoming publication.

\section{Theoretical context}

The main features of the exclusive DPE Higgs boson signal, and of the
various backgrounds are summarized below.

\subsection*{Exclusive DPE}

The first proposed  model for $pp \rightarrow p+H+p$, the Bialas-Landshoff (BL) model, 
is based on a summation of two-gluon exchange Feynman graphs coupled to Higgs production by the top quark loop. 
The non-perturbative character of diffraction at the proton vertices relies on the introduction of ``non-perturbative'' 
gluon propagators which are modeled on the description of soft total cross-sections within the additive 
constituent quark model. Reggeization is assumed in order to recover the usual parameters of the Donnachie-Landshoff 
Pomeron \cite{donnachie}. Expressions for the resulting cross-section can be found in \cite{bialas1}.

Soon after, the same model was applied to $pp \rightarrow p+q\bar{q}+p$ \cite{bialas2}. 
The computation of diffractive gluon pair production, $pp \rightarrow p+gg+p$, was performed in this 
framework very recently \cite{bzdak}.

One important aspect for the consistency of the model is the non-trivial factorization of the 
sum of all relevant diagrams as the product of a soft component by a hard elementary cross-section.
For both processes $gg \rightarrow gg$ and $gg \rightarrow
q\bar{q}$, the elementary cross-section corresponds to what would be obtained by
a separate computation imposing that the initial gluons are in the
$J_Z=0$ state. The hard $gg\to q\bar{q}$ cross-sections turn out to
be proportional to $m_{q}^{2}/s$, and hence are suppressed at high
energy. This makes the Higgs boson search in this channel theoretically attractive.

The other popular model for exclusive DPE has been developed by Khoze, Martin, Ryskin 
(KMR)~\cite{khoze}. It relies on a purely perturbative, factorized QCD
mechanism applied to 2-gluon exchange among the protons,
without reference to a reggeized Pomeron, and convoluted with the hard
sub-processes $gg\to gg, q\bar{q}$, $H$. In this context, the perturbative Sudakov form factors
are providing a sort of ``semi-hard'' cut-off which allows one to
avoid the infrared divergence in the loop integration over 
the perturbative gluon propagators. The main ingredients of this model are the so-called unintegrated off-forward 
gluon distributions in the proton, which are a source of uncertainty
\cite{lonnblad}. The hard cross-sections are computed with the $J_Z=0$
constraint on the initial gluons. Besides this aspect, 
the rapidity gap or proton survival probability, ensuring that the incoming hadrons do not re-scatter and indeed leave the 
interaction intact, have been computed and applied by the authors, using information from soft elastic scattering,
and low mass and high mass diffractive scattering \cite{khozeS2}. For a Higgs boson of 120 GeV produced at the LHC, 
the survival probability is found to be $\sim\!3\%$.

The survival probability has not been applied in the original computations by Bialas et al, and the dijet cross-sections 
are found to exceed the CDF experimental bound \cite{cdfdijet}. It has however recently been shown\footnote{This has been 
derived and tested first in the context of factorization breaking in single diffraction at HERA and the 
Tevatron~\cite{bialas3}, and later extended and generalized to double diffraction at hadron colliders\cite{bialas_pesch}.}, 
using the Good-Walker and Glauber formalisms, 
that the double Pomeron exchange contribution to central diffractive production of heavy objects has to be corrected 
for absorption, in a form determined by the elastic scattering between the incident protons. When 
applied to Higgs boson production, this leads to a strong damping factor, very comparable to the KMR factor 
\cite{alexander}. Taking this factor into account brings the dijet cross-sections in agreement with the abovementioned
experimental bound.

Monte-Carlo simulations, using {\tt DPEMC}, based on the BL model and including the rapidity-gap survival probability as 
determined above, give cross-section results compatible with the KMR model. Hence our results on the signal to background 
ratios are expected to be valid for both the gap survival corrected Bialas-Landshoff model and the KMR model.

\subsection*{Inclusive DPE and non-diffractive backgrounds}

Since the signal of interest is $pp \rightarrow p+(H\rightarrow b\bar{b})+p$, all processes involving dijets 
in the final state need to be considered as potential backgrounds. We consider them in turn.

Standard (non-diffractive) QCD dijet events constitute the most copious background. It is important in the 
early stages of the analysis (namely as a background to the first level experimental trigger), and is rejected 
requiring the detection of forward protons. These events are modeled using the {\tt PYTHIA} event generator \cite{pythia}, 
with standard QCD parameter settings.

Inclusive DPE dijet events are the following background component and are also in principle reducible, since contrarily to 
exclusive DPE, the Pomeron remnants will prevent the appearance of rapidity gaps in the central detectors. 
However, in typical LHC running conditions, a large number of interactions are present simultaneously in 
the detector, and the majority of non-diffractive events will fill the gaps left by the occasional exclusive 
DPE event. It is thus not clear whether one can expect to take benefit from this aspect of the signal.

Another way to discriminate between inclusive and exclusive DPE is to compare the dijet mass measured in the 
central detectors to the so-called missing mass, defined as the deficit between the total LHC center-of-mass energy
and the mass of the outgoing proton pair. The ratio of these quantites should be $\ssim 1$ in exclusive DPE, and 
smaller than 1 in inclusive DPE. However, the gluon density in the Pomeron has a significant component at large 
momentum fraction, and a fraction of inclusive DPE events will resemble exclusive events from this point of view. 
Inclusive DPE is thus an important background to consider. In this study, inclusive DPE dijets are simulated following 
the BPR model, with cross-sections and normalization given in \cite{us}.

The exclusive DPE dijet background has been discussed in the previous section. All DPE processes are 
simulated using {\tt DPEMC}, with settings as described in \cite{dpemc}, or with {\tt DIFFHIGGS} \footnote{This program 
is unpublished and superseded by its public version, {\tt DPEMC}.}, the program used in~\cite{us}.

\subsection*{Simulation and cross sections}

The Higgs boson events are generated using {\tt DPEMC}. Including the survival factor, the exclusive cross section 
at $\sqrt{s}$=14 TeV is found to be 2.3 fb for a Higgs boson mass of 120 GeV decaying to b quark pairs. 
We also use {\tt DPEMC} to produce the exclusive b jets. The cross section requiring jets with $p_T>25$ GeV, is 1.2 pb.
These numbers are obtained with model parameters set as in the original publications \cite{bialas1,bialas2}.

We performed two cross-checks which will be detailed in an forthcoming paper to verify the predictions of our generator.
First, we computed the cross-section for DPE dijets within the CDF acceptance, after a dijet mass fraction cut at 0.8,
as it is done by the CDF experiment. We found a cross section of about 0.16~nb, well below the experimental 
bound of 3.7~nb. The other test was to check the suppression factor of exclusive b jets with respect to all other 
jets: we find 
a b-quark dijet cross section of about 2.1 pb after a jet $p_T$ cut of 25 GeV, and 6$\times$10$^3$ pb for all 
quark and gluon jets, the total quark contribution being 2.3 pb. This corresponds to the expected suppression of quark
pair production in exclusive DPE.

The inclusive background has been generated using the {\tt DIFFHIGGS} Monte Carlo. In order to limit the size of the 
simulated samples, we require jets with $p_T$ greater than 25 GeV,and a dijet mass greater than 75 GeV. The protons 
are also required to fall within the forward detector acceptance (see next section), and the missing mass is required 
to be between 100 and 170 GeV. The resulting inclusive DPE dijet cross-section is 22 pb.

\section{Experimental context}

This section summarizes the characteristics of the LHC detectors
relevant to this study.

\subsection*{The central detector}

The analysis below relies on a fast simulation of the CMS detector at the LHC. The same study could 
be performed using the ATLAS detector simulation, when one would expect similar results. The relevant
detector characteristics are briefly recalled below.

The calorimetric coverage of the LHC experiments ranges up to a pseudorapidity of $|\eta|\ssim 5$. The region devoted
to precision measurements lies within $|\eta|\leq 3$, with a typical resolution on jet energy measurement of $\ssim\!50\%/\sqrt{E}$,
where $E$ is in GeV, and a granularity in pseudorapidity and azimuth of $\Delta\eta\times\Delta\Phi \ssim 0.1\times 0.1$. 
For dijets, the mass resolution at $M_{JJ}\ssim 100$ GeV is about 10\%.
The extension in the forward region $3<|\eta|<5$ allows a precise measurement of the missing transverse energy,
and can be used to select rapidity gaps by vetoing activity in this region (in the absence of pile-up).

The identification of b-quarks is done by detecting the decay vertices of B mesons. This is done by searching for displaced vertices,
or for charged particle tracks with a large impact parameter with respect to the interaction point. The light quark or gluon jet
rejection depends on the chosen b-quark selection efficiency; typically, one expects a rejection factor of 100 for a selection efficiency
of 60\%. For a Higgs boson decaying to b-quark pairs, the efficiency is $\ssim 35\%$, and the non-b dijet background is rejected by a 
factor $10^4$.

\subsection*{The forward detector}
A possible experimental setup for forward proton detection is described in 
\cite{helsinki}. We will only briefly recall its features here, and will 
concentrate on its acceptance and resolution.

Protons diffracted at very low angles, or with a small momentum loss, are detected at large distances 
from the interaction point when, following the machine optics, they have 
sufficiently deviated from the nominal beam. 

In exclusive DPE, the mass of the central heavy object is given by $M^2 = \xi_1\xi_2 s$, where $\xi_i$ are 
the proton fractional momentum losses, and $s$ is the total center-of-mass 
energy. In order to reconstruct objects with mass 100-150 GeV in this 
way, the acceptance sould be large down to $\xi$ values as low as a few 
$10^{-3}$. The missing mass resolution directly depends on the resolution on 
$\xi$, and should not exceed a few percent if a significant improvement compared to the dijet mass resolution
is desired \cite{rosto}.

These goals are achieved in \cite{helsinki} assuming three detector stations, located at $\sim 210$ m, $\sim 
308$ m, and $\sim 420$ m from the interaction point. According to the 
currently foreseen LHC machine parameters, protons with a momentum loss of a few $10^{-3}$
will be sufficiently separated from the beam envelope only 
after having traveled such large distances. The $\xi$ acceptance and 
resolution have been derived for each device using a complete simulation
of the LHC beam parameters. The combined $\xi$ acceptance is $\sim 100\%$ 
for $\xi$ ranging from $0.002$ to $0.1$. The acceptance limit of the device closest to the interaction point
is $\xi > \xi_{min}=$0.02.

The present analysis does not assume any particular value for the $\xi$ resolution. 
Instead, for the sake of generality, results are presented as a function of 
the final missing mass resolution, so that the search performance of any 
given setup can be read off directly.

\section{Sensitivity to the Standard Model Higgs Boson}

This section gives an overview of the selection procedure of exclusive DPE Higgs boson events.
We consider trigger strategies relying on rapidity gaps and forward proton detection, their domain of application 
and their limitations. The analysis is then described, and the results follow.

\subsection*{Triggering with forward protons}

Let us first discuss possible trigger strategies for this channel. The
dijet cross-section at the LHC is orders of magnitude too large to allow triggering 
on the jets themselves, so benefit must be taken from the specifities of DPE.

If the needed $\xi$ acceptance can be obtained for detectors close enough to the interaction point,
requiring at least one detected proton at the first level trigger eliminates all non-diffractive
dijet events and solves the triggering problem. The maximum allowed distance is about 200-250~m, a number given by the time needed
for a proton to fly from the interaction point to the forward detector, for the detector signal to travel back,
and for the trigger decision to be made, within the allowed first level trigger latency. This latency is about 1.8~$\mu$s 
for the ATLAS detector; CMS disposes of about 3~$\mu$s.

\begin{figure}[htbp]
\begin{center}
\caption{Proton momentum loss distribution, for an exclusive DPE Higgs
boson signal ($m_H = 120$ GeV). The forward proton acceptance is shown
for the whole detection system, and for the device closest to the
interaction point.}
\epsfig{file=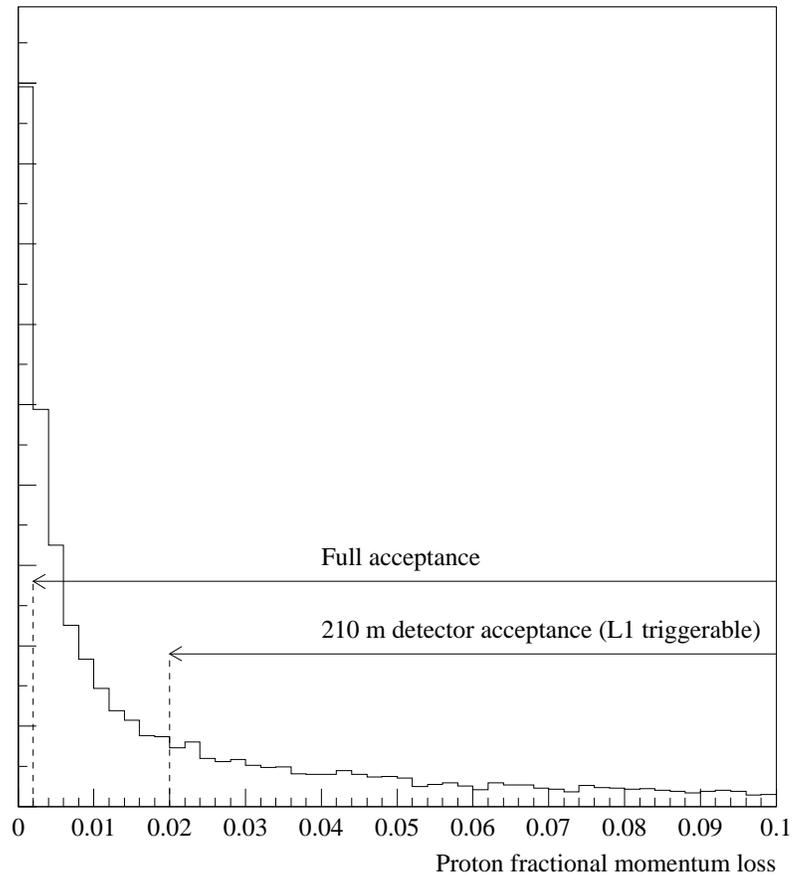,width=.75\textwidth}
\label{figpot1}
\end{center}
\end{figure}

Figure~\ref{figpot1} shows the proton $\xi$ distribution for a Higgs boson mass of 120~GeV.
Given the $\xi$ acceptance of the closest detector
($\xi_{min}=0.02$), requiring one proton to be detected at the first level trigger has an
acceptance of about 66\%. If one proton satisfies $\xi > 0.02$, the second one
has much smaller momentum loss and can be detected in the large
distance devices. Requiring the detection of both protons in the short
distance devices has acceptance only above $m_H=\sqrt{\xi_{min}^{2}s}=280$~GeV.

Requiring in addition two jets with tranverse momenta of at least 40 and 30 GeV gives 
a first level trigger rate of about 80 Hz at a luminosity L~$= 2 \times 10^{33}$~cm$^{-2}$~s$^{-1}$, and 400 Hz at L~$=
10^{34}$~cm$^{-2}$~s$^{-1}$. These numbers correspond to the low and
high luminosity running scenarios at the LHC.
If the detection of the second proton is required at a higher trigger
level, and a cut on the missing mass is added (for instance $80 < M_{miss} < 250$ GeV, 
where $M_{miss}^2=\xi_1 \xi_2 s$), the final trigger rate is less than 0.2 Hz (1 Hz) at low 
(high) luminosity.
\subsection*{Triggering with rapidity gaps}

If the strategy proposed in the previous section is insufficient (i.e,
if the forward detector signal arrives beyond the latency limit, or if the
quoted single proton detection efficiency is too low), the trigger has
to rely on central detector signals.

The first level trigger rate requiring two jets with $p_T >$ 40 and 30 GeV, and a 
dijet mass greater than 80 GeV, is about 10 kHz at low luminosity and 100 kHz at high luminosity. 

\begin{figure}[htbp]
\begin{center}
\caption{Total transverse energy distributions in the forward calorimeters
($3 < |\eta| < 5$), for relevant non-diffractive and DPE processes.}
\epsfig{file=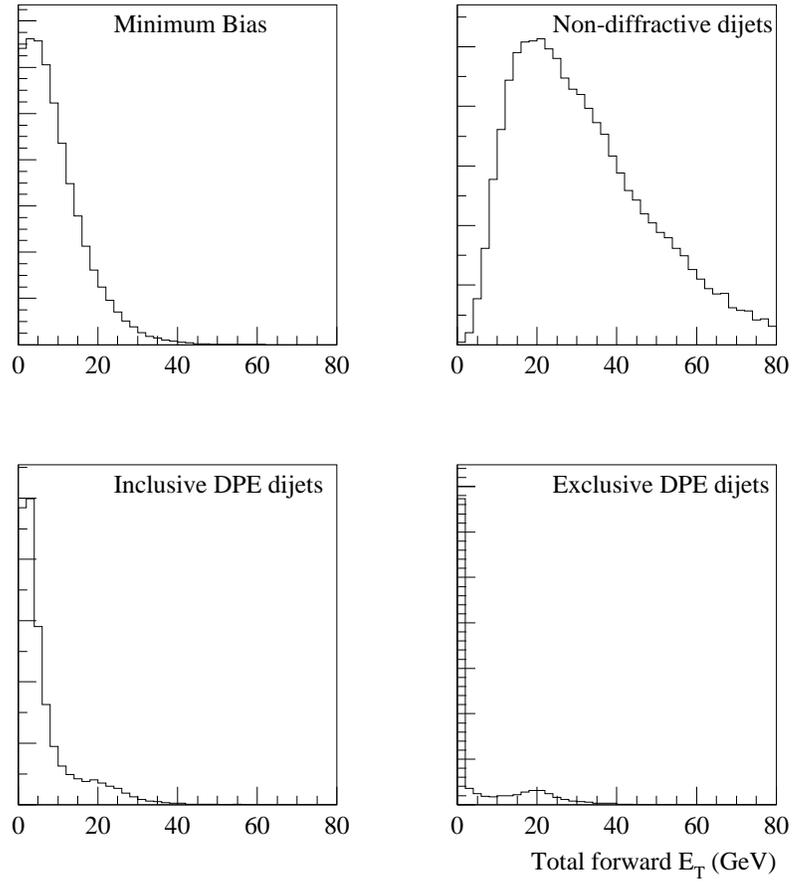,width=.75\textwidth}
\label{figgap2}
\end{center}
\end{figure}

It is in principle possible to reduce this rate at Level 1 by
requiring rapidity gaps between the protons and the jets. As Figure~\ref{figgap2}
shows, requiring the absence of activity in the forward calorimeters
(by requiring the total transverse energy in this region to be low)
effectively selects DPE events against non-diffractive dijet
events. So this appears to be a simple and promising strategy.

However, at high luminosity, some twenty interactions occur
simultaneously and overlap in the detector. Figure~\ref{figgap2} shows
again that, even if an exclusive DPE event has no forward calorimetric activity,
the superimposition of minimum bias events washes out this feature,
and spoils the discrimination between diffractive and
non-difractive events.

To profit from diffractive signatures in the central detectors, it
thus appears desirable to run at lower luminosity, in order to
maximize the rate of single interaction collisions. In fact, one can
express the probability to observe exactly one interaction of low
cross-section (and no overlapping minimum bias events) as follows:

$$\mathrm{P \propto L \exp\frac{-\sigma_{mb} L}{f}}$$

\noindent where L is the luminosity, $\mathrm{\sigma_{mb}}$ is the minimum bias
cross-section, taken to be 55 mb, and f is the crossing frequency,
which is 40 MHz at the LHC.

\begin{figure}[htbp]
\begin{center}
\caption{Evolution of the probability to observe exactly one
interaction during an LHC bunch crossing, as a function of the machine
luminosity.}
\epsfig{file=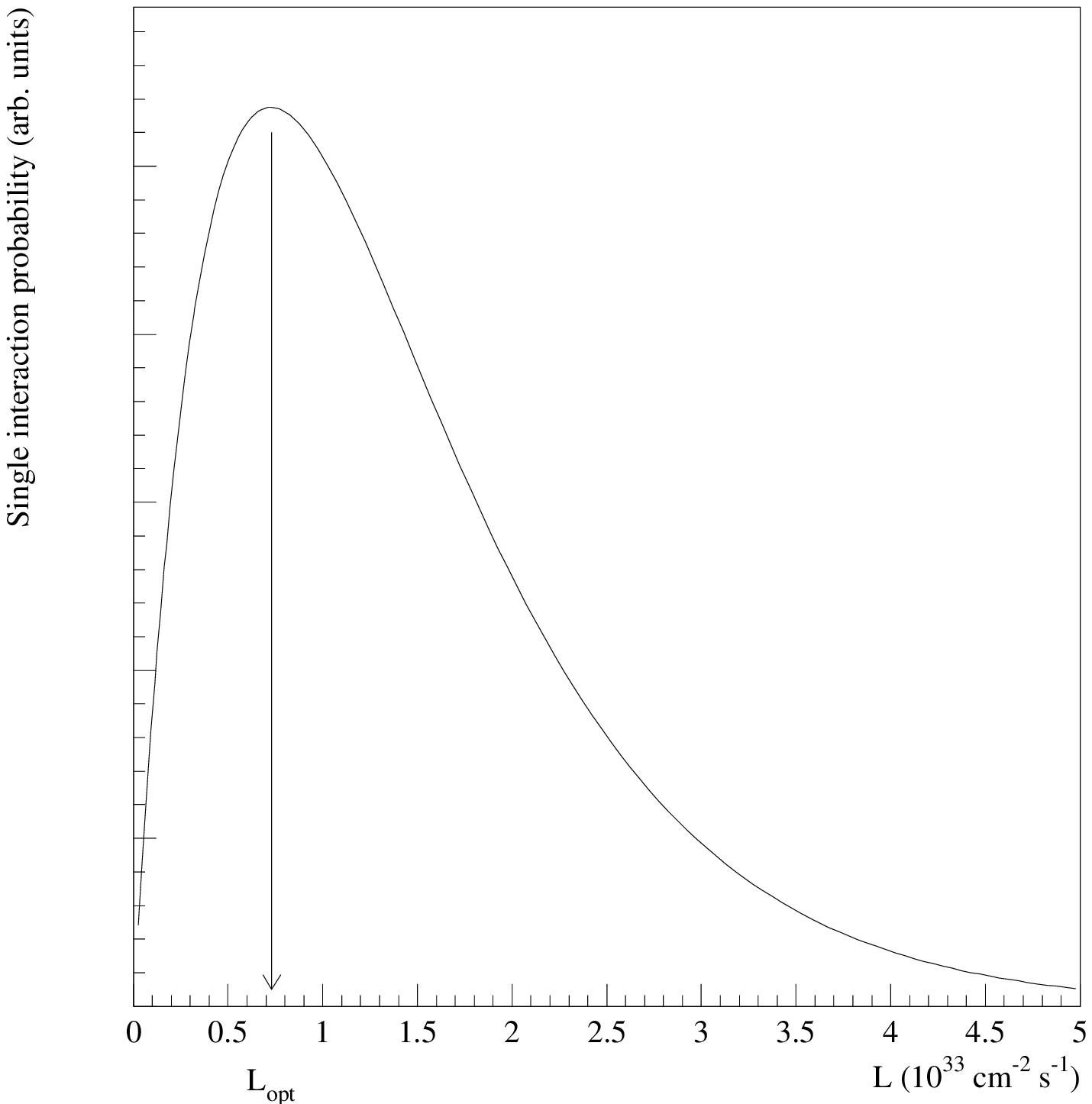,width=.75\textwidth}
\label{figgap1}
\end{center}
\end{figure}

The behaviour of this function is displayed in
Figure~\ref{figgap1}. The value of L maximizing the single interaction
rate is $\mathrm{L_{opt}} = 7.3 \times 10^{32}$~cm$^{-2}$~s$^{-1}$. Note also that at this
``optimal'' luminosity, the average number of overlapping events is
still $\bar{n} = \mathrm{\sigma_{mb} L_{opt} / f} = 1$, so that that
the fraction of events without overlaps is $e^{-1}=0.37$. One can thus
define an effective luminosity as $\mathrm{L_{eff} = L_{opt} \times
e^{-1}} = 2.7 \times 10^{32}$~cm$^{-2}$~s$^{-1}$, which determines the
counting rate of clean DPE events without pile-up. Obviously, rare
signals accumulate very slowly under these conditions.

We do not exclude that clever ways can be found that allow to
distinguish DPE events from non-diffractive dijets in the presence of
pile-up. But this requires excellent detector understanding and
knowledge of minimum bias processes. This study will be performed in a forthcoming publication.

At a higher trigger level, the information from forward detectors can be used, and the final rates 
will be at the same level as before. But we stress that it is crucial
for the experiments to maintain a manageable trigger rate at the first
level. Considering the available bandwidth (75 to 100 kHz for Atlas,
and a similar number for CMS), and the concurrence of other important
trigger channels, a few hundred Hz appears to be a maximum.
 
\subsection*{Analysis}

This section summarizes the cuts applied in the remaining part of the analysis.
As said before, both diffracted protons are required to be detected in roman pot detectors. The central mass 
is reconstructed using the measurement of $\xi_1$ and $\xi_2$ given
by the forward detectors, giving $M_{miss}= (\xi_1 \xi_2 s)^{1/2}$. The resolution
on the central mass is thus directly dependent on the leading proton measurement resolution.
As mentioned before, we choose to study the signal to background ratio as a function 
of the missing mass resolution, by varying this parameter directly.

\begin{figure}[htbp]
\begin{center}
\caption{Mass fraction distribution for inclusive and exclusive DPE dijets events.}
\epsfig{file=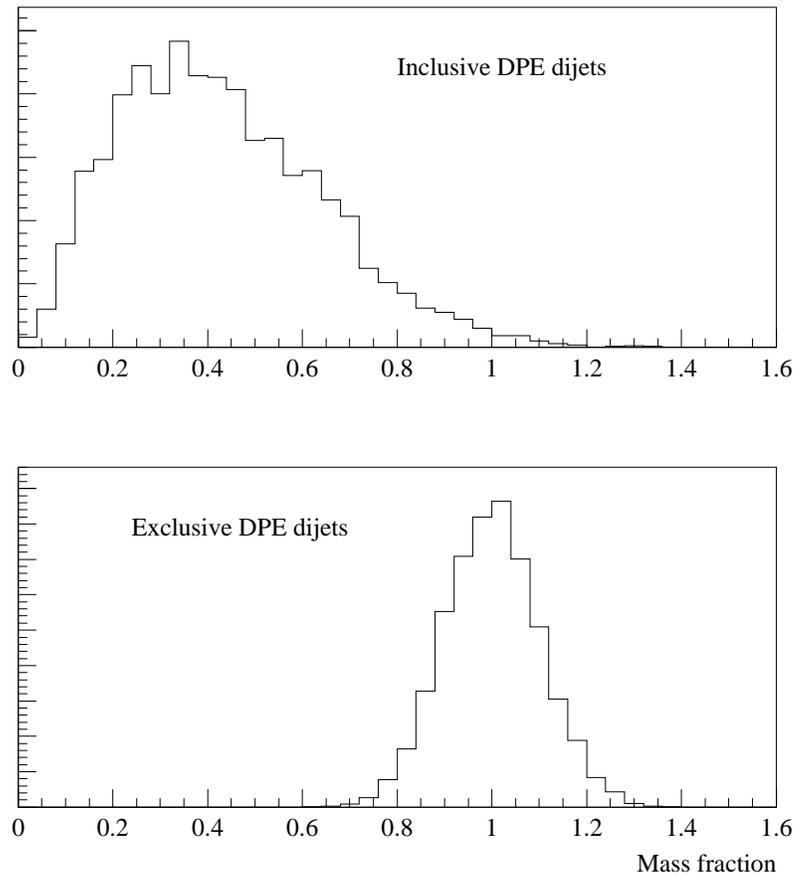,width=.75\textwidth}
\label{figana1}
\end{center}
\end{figure}

The other cuts are based on detecting well measured, high $p_T$ $b\bar{b}$ events. For this,
we use a fast simulation of the CMS detector (the ATLAS detector simulation will produce very similar results). 
We first require the presence of two jets with $p_{T1} >$ 45 GeV, $p_{T2} >$ 30 GeV. The difference in 
azimuth between the two jets should be $170 < \Delta \Phi < 190$ degrees, asking the jets to be back-to-back. 
Both jets are required to be central, $|\eta| <$  2.5, with the difference
in rapidity of both jets satisfying $| \Delta \eta | < 0.8$. We also apply
a cut on the ratio of the dijet mass to the total mass of all jets
measured in the calorimeters, $M_{JJ}/M_{all} >$ 0.75. The ratio of the dijet mass 
to the missing mass should verify $M_{JJ} / (\xi_1 \xi_2 s)^{1/2}
>$~0.8. As can be seen on Figure~\ref{figana1}, the mass fraction
distribution for exclusive events has a spread of about 10\%,
dominated the dijet mass resolution as expected. 

An additional cut requires a positive $b$ tagging of the jets, eliminating all non-b dijet background, with
the efficiency on b-quark dijets quoted above.

The last important cut requires that all the available Pomeron-Pomeron collision energy is used to produce 
the Higgs boson. Such a topology could be selected by
requiring the dijet mass to be close to the total mass measured in the calorimeters (i.e summing over all calorimeter
cells, rather than over all jets as done above). Such a selection clearly needs to be controlled accurately and would 
need a more complete simulation of the calorimeter response, notably including a detailed noise and pile-up simulation.
The present study emulates this cut by requiring the Pomeron momentum fraction involved in the hard process to be greater
than 95\%. This cut appears crucial in eliminating the inclusive DPE background.
\subsection*{Results}

\begin{figure}[htbp]
\begin{center}
\caption{Standard Model Higgs boson signal to background ratio as a function of the resolution on 
	the missing mass, in GeV. This figure assumes a Higgs
	boson mass of 120 GeV. }
\epsfig{file=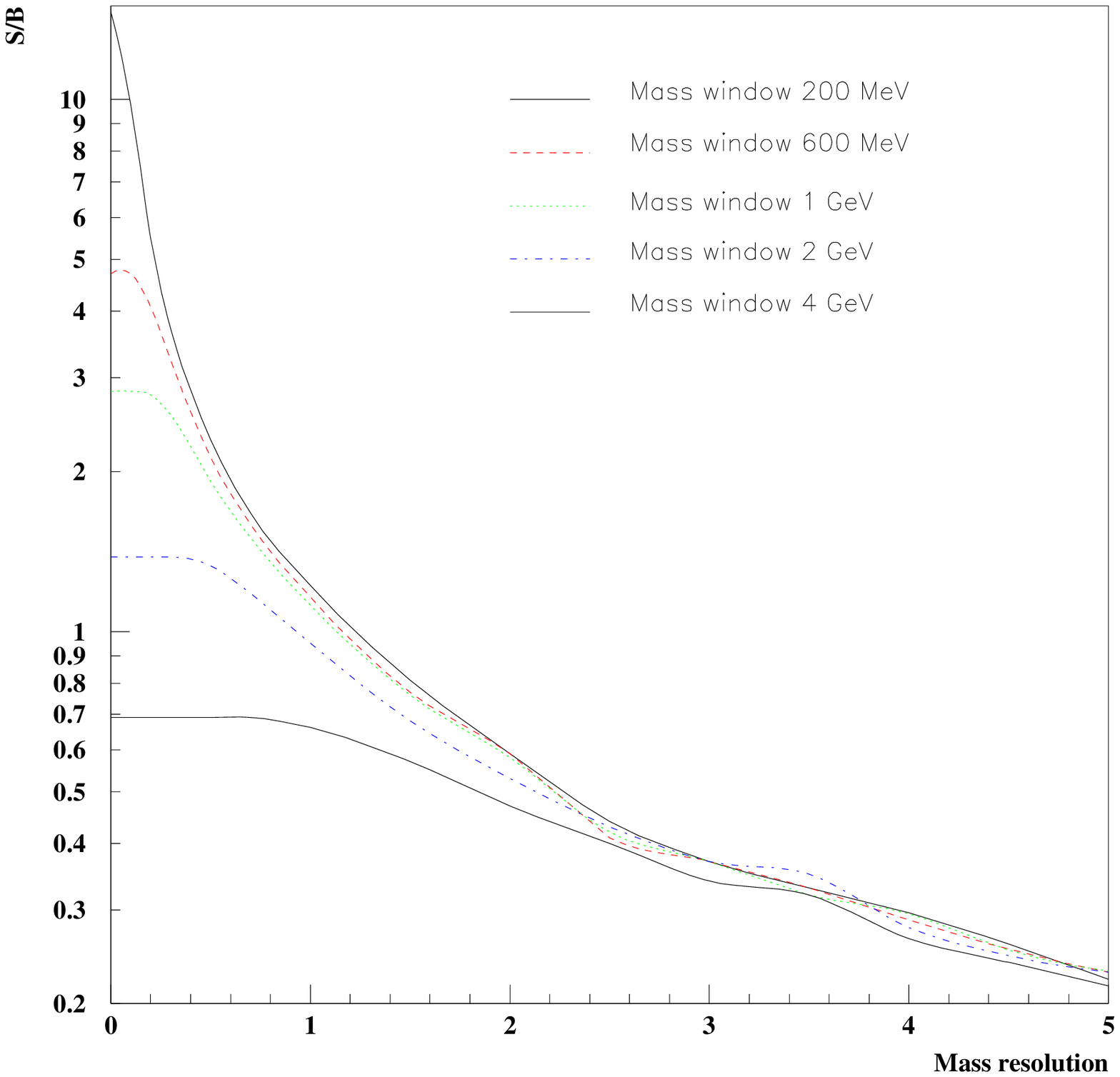,width=.75\textwidth}
\label{figres1}
\end{center}
\end{figure}

\begin{figure}[htbp]
\begin{center}
\caption{Background (fitted by an exponential) and signal superimposed, 
	for an example mass resolution of 2.5 GeV (arbitrary normalization). This figure assumes a Higgs
	boson mass of 120 GeV.}
\epsfig{file=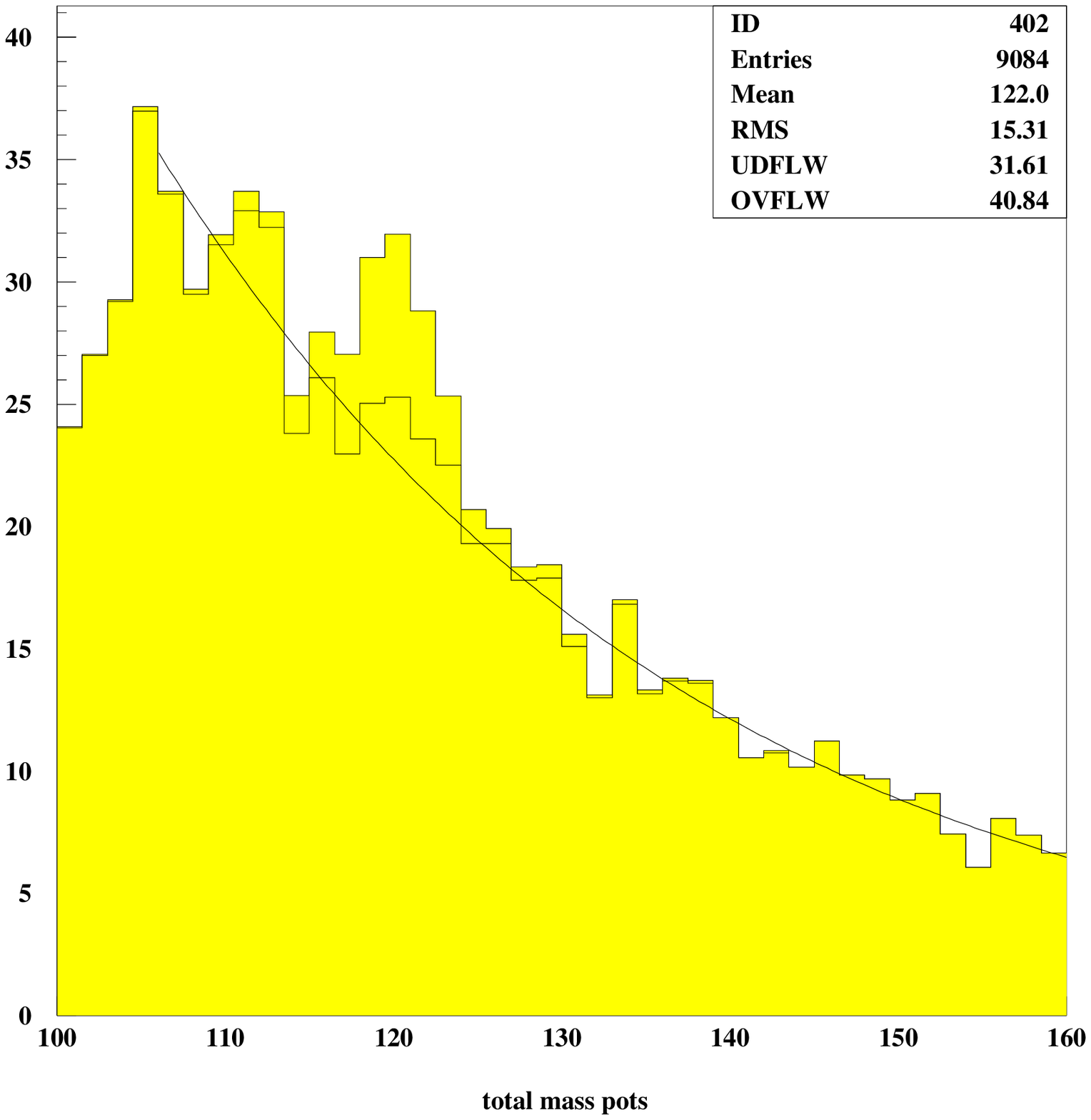,width=.75\textwidth}
\label{figres2}
\end{center}
\end{figure}

Results are given in Figure~\ref{figres1} for a Higgs mass of 120 GeV, in terms of the signal to background 
ratio S/B, as a function of the Higgs boson mass resolution. The background and overlayed signal is shown
in Figure~\ref{figres2}, for an example mass resolution of 2.5 GeV.

In order to obtain an S/B of 3 (resp. 1, 0.5), a mass resolution of about
0.3 GeV (resp. 1.2, 2.3 GeV) is needed. The forward detector design of \cite{helsinki} 
claims a resolution of about 2.-2.5 GeV, which leads to a S/B of about 0.4-0.6. Improvements in this design
would increase the S/B ratio as indicated on the figure.

For 100 fb$^{-1}$, one expects of the order of 20 signal events, when using a mass resolution  of about 2.5 GeV and within 
a mass window of 4 GeV. As usual, this number is enhanced by a large factor if one considers supersymmetric Higgs boson 
production with favorable Higgs or squark field mixing parameters.

Finally, let us note that the background increases by a factor 5 if the last cut of the analysis is not
applied (see previous section), due to contamination by inclusive events. As a result, S/B $\sim$ 0.1.

\subsection*{Comparison with the KMR estimate}

Our result can be compared to the phenomenological result of
\cite{dkmro}, where experimental issues were addressed within  
the KMR framework. For a missing mass resolution of $\sim$1 GeV, we
have obtained S/B$\sim$1, where the KMR collaboration  
finds S/B$\sim$3. Although our analysis relies on a more detailed
experimental simulation, the reason for the difference is elsewhere. 

In \cite{dkmro}, the background is integrated over a mass window of
1~GeV, assuming that 100\% of the signal lies inside this window. This
is the case only if the mass resolution is significantly smaller   
than 1~GeV, and typically of order 250-300~MeV.

So assuming the result of \cite{dkmro} is given for a gaussian mass
resolution of 1 GeV either underestimates the background by a factor
$\sim$3, or overestimates the signal by the same factor. Taking this
factor into account, and once again assuming that trigger rates and
contamination by inclusive DPE can be kept under control, brings the
KMR estimate to agree with our Monte-Carlo simulation.

\section{Summary}

We have performed a Monte-Carlo simulation of the exclusive DPE Standard Model Higgs boson search, accounting 
for the signal, backgrounds, and detector effects in a realistic way.

We stressed that the trigger strategy for such a signal is straightforward, provided the forward detector signals 
arrive early enough. This strongly limits the allowed distance between the forward detectors and the interaction point.
The $\xi$ acceptance criteria are contradictory to the previous condition, and prefer larger distances. If no compromise
can be found, the trigger has to rely on the central detectors only. Rapidity gaps can provide an efficient trigger signal,
but only at low luminosity, which means that the signal accumulates slowly.

The selection of exclusive DPE events is difficult because of the contamination by inclusive events. It is found that the 
``quasi-exclusive'' tail of inclusive DPE (with a dijet to missing mass ratio larger than 0.8) is hard to eliminate, and requires 
selections that are very sensitive to detector effects. Further investigation in this direction is needed.

If the above difficulties can be overcome, i.e if it is possible to trigger on DPE events efficiently, and select exclusive 
DPE with high purity, then the signal to background ratio is a factor three smaller than predicted elsewhere. Quantitatively, a 
missing mass resolution of 1 GeV implies S/B of order 1; to obtain S/B of order 3, a resolution of a few hundred MeV is required.

\end{document}